\documentclass[12pt,twoside]{article}
\usepackage{wrapfig}
\usepackage{graphicx}
\usepackage{epsfig}
\usepackage{cmp2e}
\font\bba=msbm10 scaled 1080
\font\bbb=msbm8 
\font\bbc=msbm6 
\newfam\bbfam
\textfont\bbfam=\bba \scriptfont\bbfam=\bbb
\scriptscriptfont\bbfam=\bbc
\def\bb{\fam\bbfam\bba}

\def\Z{{\bb Z}}

\title[The collective variables representation of simple fluids ]%
{The collective variables representation of 
simple fluids from the point of view of statistical field theory.
\thanks{This paper is dedicated to Professor Ihor Yukhnovskii on the
occasion of his $80^{th}$ birthday.}}
\author[J.-M.~Caillol, O.~Patsahan, and I.~ Mryglod]{Jean-Michel Caillol \refaddr{label1},
        Oksana Patsahan \refaddr{label2}, 
        and Ihor Mryglod \refaddr{label2}
          }
\addresses{
\addr{label1} Laboratoire de Physique th\'eorique \\
CNRS UMR 8627, B\^at. 210 \\
Universit\'e de Paris-Sud \\
91405 Orsay Cedex France
\addr{label2} Institute for Condensed Matter Physics \\
of the National Academy of Sciences of Ukraine\\
1 Svientsitskii Str.\\79011 Lviv Ukraine
}
\begin{document}

\maketitle

\begin{abstract}
The collective variable representation (CV) of classical statistical systems such as simple liquids has been intensively developed by the Ukrainian school after seminal works by Prof. Ihor Yukhnovskii. The basis and the structure of the CV representation are reexamined here from the point of view of statistical field theory and compared with another exact statistical field representation of liquids based upon a Hubbard-Stratonovich transform. We derive a two-loop expansion for the grand potential and free energy  of a simple fluid in both version of the theory. The results obtained by the two approaches  are shown  to coincide at each order of the loop expansion. The one-loop results are identical to those obtained in the framework of the random phase approximation of the theory of liquids. However, at the second-loop level,  new expressions for the pressure and the free energy are obtained, yielding a new type of approximation.
\pacs Up to six PACS numbers
\end{abstract}

\section{Introduction}

Functional methods in modern statistical physics represent one of the most powerful tools for the study of both equilibrium and dynamical properties (see, e.g. \cite{Zinn,Orland}). In many cases the partition function of simple
models (Ising and  Heisenberg spins or classical particles in interaction) can be re-expressed as a functional integral after performing a Hubbard-Stratonovich transformation, a simple device 
proposed in the 50ies  \cite{Hubbard1,Strato} . Nearly at the same time  another method - the so-called collective variables (CV) method - that allows in a explicit way to construct a functional representation for many-particle interacting systems was developed \cite{Zubar,Yuk1} and applied for the description of charged particle systems, in particular, to the calculation of the configurational integral of the Coulomb systems. The idea of this method is based on: (i) the concept of collective coordinates being appropriate for the physics of system considered (see, for instance, \cite{Bohm-Pines}), and (ii) an integral giving an exact functional representation for the configurational Boltzmann factor. Later the CV methods was successfully developed for the description of classical many-particle systems \cite{Yuk-Hol} and for the theory of second order phase transitions \cite{Yuk}. The method has a wide range of application and,  to paraphrase Pr. Yukhnovskii in his book \cite{Yuk} " The collective variables is a common name of a class of variables which are specific for each kind of physical system. For magnetic systems CV are variables corresponding to the modes of vibrations of spin moments, for ferroelectrics they are cluster vibrational modes, for systems of charged particles they are associated with the modes of generalized charge vibrations, for binary alloys they are modes of one-particle distribution function, the CV of the liquid-gas system are modes of some deviations from the critical density and so on." 

One of the goals of this paper is to reconsider the CV method from the point of view of  statistical field theory  and to
compare the results obtained with those found recently by one of us by means of the KSSHE
(Kac-Siegert-Stratonovich-Hubbard-Edwards) theory \cite{Cai-Mol}.

We formulate the method of CV in real space and consider a one-component continuous model consisting of hard spheres interacting through additive pair potentials. The expression for the functional of the grand partition function is derived and the CV action that depends upon two scalar fields - field $\rho$ connected to the number density of particles and field $\omega$ conjugate to $\rho$ and connected to the density of energy - is calculated. We study correlations between these fields as well as their relations to the density and energy correlations of the fluid. The grand partition function of the model is evaluated in a systematic way using a well-known method of statistical field theory, namely the so-called loop expansion.
It consists in expanding functionally the action $\mathcal{H}$ around a saddle point, so that the lowest order (zero loop)
approximation defines the mean-field (MF) level of the theory and the first order loop expressions correspond to the random phase approximation (RPA). Recently \cite{Cai-Mol} this technique was applied to the action obtained within the framework of the KSSHE theory. In this paper we perform a two-loop expansion of the pressure and the free energy of the homogeneous fluid which yields a new type of approximation which we plan to test in our future work.

The paper is organized as follows. In Section~(\ref{Proleg}), starting from the Hamiltonian, we introduce the two different functional representations of the grand partition function based on the KSSHE and CV methods. Here we also enter several types of statistical field averages that are important in the further part of the
paper. In Section~(\ref{Corre}) we introduce the CV and KSSHE field correlation functions, establish links between them as well as their relation to the density correlation functions of the fluid.
The MF level of the KSSHE and CV field theories is formulated in
Section~(\ref{MF}). Section~ (\ref{loop}) is devoted to the loop expansion of the grand
potential. The pressure and the free energy of the homogeneous
fluid are obtained in the two-loop approximation in Section~6. We
conclude with some final remarks in Section~(\ref{conclu}).
\section{Prolegomena}
\label{Proleg}
\subsection{The model}
\label{model}
We consider  the case of a simple three dimensional  fluid made  of identical hard spheres of diameter $\sigma$ with
additional isotropic pair interactions $v(r_{ij})$ ($r_{ij}=| x_i -x_j |$, $x_i$ is the position of particle "$i$"). Since $v(r)$ is an arbitrary function of $r$  in the core, i.e. for $r \leq \sigma$, we will assume that $v(r)$ has been regularized in
such a way that  its Fourier transform $\widetilde{v}_{q}$ is a well behaved function of $q$ and that $v(0)$ is a finite quantity. We denote by $\Omega$ the domain of volume $V$ occupied by the molecules of the fluid. The fluid is at equilibrium in the grand canonical (GC) ensemble, $\beta=1/k_{\mathrm{B}}T$ is the inverse temperature ($k_{\mathrm{B}}$ Boltzmann constant), and $\mu$  the chemical potential. In addition the particles are subject to an external potential $\psi(x)$ and we will denote by $\nu(x)=\beta (\mu-\psi(x))$ the dimensionless local chemical potential. We will stick to notations usually adopted in standard textbooks devoted to the theory of liquids (see e.g. \cite{Hansen}) and will  denote by $w(r)=-\beta v(r)$  \emph{minus } the dimensionless pair interaction. Quite arbitrarily we will say that the interaction is \emph{attractive} if the Fourier transform $\widetilde{w}(q)$ is\emph{ positive} for all $q$; in the converse case it  will be said repulsive.

In a given GC configuration $\mathcal{C}=(N;x_1 \ldots x_N)$ the microscopic density of particles reads
\begin{equation}
\widehat{\rho}(x|\mathcal{C}) =
\sum_{i=1}^{N} \delta^{(3)}(x-x_i) \; ,
\end{equation}
and the GC partition function $\Xi\left[ \nu \right] $ can thus be written as
\begin{eqnarray}
\label{csi}\Xi\left[ \nu \right] &=&
 \mathrm{Tr}\left[ \; \exp\left( -\beta
V_{\mathrm{HS}}(\mathcal{C}) +\frac{1}{2} \left\langle
\widehat{\rho}|w \vert\widehat{\rho} \right\rangle +
\left\langle  \overline{\nu}|\widehat{\rho} \right\rangle
\right)\right]  \nonumber  \; ,\\ 
 \mathrm{Tr}\left[  \ldots \right] &=&
 \sum_{N=0}^{\infty}
\frac{1}{N!} \int_{\Omega}d1 \ldots dn \ldots \; ,
\end{eqnarray}
where $i \equiv x_i $ and $di\equiv d^{3}x_i$. For a given volume $V$, $\Xi\left[ \nu \right]$ is a function of $\beta$ and a convex functional of the local chemical potential $\nu(x)$  \cite{dft} which we have strengthened by using a bracket.
In equation\ (\ref{csi}), $\exp(-\beta V_{\mathrm{HS}}(\mathcal{C}))$ denotes the hard sphere contribution to the Boltzmann factor  and $\overline{\nu}=\nu+\nu_S$ where $\nu_S= - w(0)/2$ is $\beta$ times   the  self-energy of a
particle. It follows from our assumptions  on $w(r)$  that  $\nu_S$ is a finite quantity which depends however on the regularization of the potential in the core. In the right hand side of equation\ (\ref{csi}) we have introduced convenient  brac-kets notations
\begin{eqnarray}
\left\langle  \overline{\nu}|\widehat{\rho} \right\rangle
 &\equiv&  \int_{\Omega} d1 \; \overline{\nu}(1)\widehat{\rho}(1) \\
\left< \widehat{\rho} |w \vert\widehat{\rho} \right> & \equiv &
\int_{\Omega} d1 d2\;
\widehat{\rho}(1|\mathcal{C})
 w(1,2)  \widehat{\rho}(2|\mathcal{C}) \; .
\end{eqnarray}
We show now how to re-express $\Xi\left[ \nu \right]$ as a functional integral.
\subsection{The Hubbard-Stratonovitch transform}
\label{HS}
The Hubbard-Stratonovich transformation \cite{Hubbard1,Strato} which was proposed in the 50ies is based on simple properties of Gaussian integrals.
In this framework the GC partition function $\Xi\left[ \nu \right]$ can be formally re-expressed as a functional integral
\cite{Hubbard1,Strato,Kac,Siegert,Hubbard2,Edwards}. Moreover
 $\Xi\left[ \nu \right]$, when re-written in this manner, can be interpreted as the GC partition function of a fluid
of bare hard spheres in the presence of an external  random Gaussian field $\varphi$  with a covariance given by the pair potential \cite{Cai-Mol,Wiegel,Cai-JSP}. It will be convenient to distinguish the  case of attractive interactions ($\widetilde{w}(q)>0$) from that of repulsive ones ($\widetilde{w}(q)>0$).
\begin{itemize}
\item{  i)}  Attractive  interactions ($\widetilde{w}(q)>0$)

We start from the well-known property of Gaussian integrals
\begin{eqnarray}
\label{toto}
\exp\left( \frac{1}{2} \left\langle \widehat{\rho} | w |   \widehat{\rho} \right\rangle  \right)&=&
\frac{\int \mathcal{D} \varphi \; \exp \left( - \frac{1}{2}
\left\langle \varphi | w^{-1} | \varphi \right\rangle + \left\langle \widehat{\rho} | \varphi  \right\rangle \right) }
{\mathcal{N}_{w} \equiv   \int \mathcal{D} \varphi \; \exp \left( - \frac{1}{2}
\left\langle \varphi | w^{-1} | \varphi \right\rangle \right) } \; , \nonumber \\
&\equiv &
\left\langle \exp
\left( \left\langle \widehat{\rho} | \varphi  \right\rangle \right) 
\right\rangle_{w} \; ,
\end{eqnarray}
where $\varphi$ is a real random field and the inverse $w^{-1}$ must be ununderstood in the operator sense, i.e.
$\int_{\Omega}d3 \; w(1,3)w^{-1}(3,2) = \delta(1,2) $. The functional integrals which enter the above
equation can be given a precise meaning in the case where the domain $\Omega$ is a cube of side $L$ with periodic boundary conditions (PBC) which will be implicitly assumed henceforth. Additional technical  details are to be found   in the  appendix. We now insert equation~(\ref{toto}) in the definition~(\ref{csi}) of the GC partition function $\Xi\left[ \nu \right]$  which yields
\begin{eqnarray}
\label{attractive}
\Xi\left[ \nu \right] &=& \left\langle \mathrm{Tr}\left[ \; \exp\left( -\beta
V_{\mathrm{HS}}(\mathcal{C}) +
\left\langle  \overline{\nu} + \varphi  |\widehat{\rho} \right\rangle
\right)\right]\right\rangle_{w}  \; ,\nonumber \\
 &=& \left\langle \Xi_{\mathrm{HS}}\left[ \overline{\nu} + \varphi\right]
 \right\rangle_{w} \; .
\end{eqnarray}
$\Xi\left[ \nu \right]$, when re-written in this manner, can thus be  re-interpreted as the GC partition function of a fluid
of bare hard spheres in the presence of a  random Gaussian field $\varphi$  with a covariance given by the pair potential \cite{Cai-Mol,Wiegel,Cai-JSP}. 
\item{ ii)} Repulsive  interactions ($\widetilde{w}(q)<0$)

This time we make use of 
\begin{eqnarray}
\label{totoi}
\exp\left( -\frac{1}{2} \left\langle \widehat{\rho} | w |   \widehat{\rho} \right\rangle  \right)&=&
\frac{\int \mathcal{D} \varphi \; \exp \left(  \frac{1}{2}
\left\langle \varphi | w^{-1} | \varphi \right\rangle +\mathrm{i}   \left\langle \widehat{\rho} |   \varphi  \right\rangle \right) }
{\mathcal{N}_{-w} \equiv   \int \mathcal{D} \varphi \; \exp \left( \frac{1}{2}
\left\langle \varphi | w^{-1} | \varphi \right\rangle \right) } \; , \nonumber \\
&\equiv &
\left\langle \exp
\left( \mathrm{i}  \left\langle \widehat{\rho} | \varphi  \right\rangle \right) 
\right\rangle_{w} \; ,
\end{eqnarray}
which yields 
\begin{equation}
\label{repulsive}
\Xi\left[ \nu \right] =
 \left\langle \Xi_{\mathrm{HS}}\left[ \overline{\nu} + i \varphi\right]
 \right\rangle_{(-w)} \; .
\end{equation}
Note that since $\varphi$ is a real scalar field  $\Xi_{\mathrm{HS}}\left[ \overline{\nu} + i \varphi\right]$ has to be evaluated for imaginary chemical potentials which may cause some troubles since $\log \Xi_{\mathrm{HS}}\left[ \nu\right]  $ has singularities (i.e. branch cuts) in the complex plane \cite{Lambert}.
In the repulsive case, the hard core part of the interaction is in fact  not compulsory to ensure
the existence of a thermodynamic limit \cite{Ruelle} and the reference system can be
chosen as the ideal gas \cite{Lambert,Efimov}.
\end{itemize}
Some comments are in order. Firstly, equations~(\ref{attractive}) and~(\ref{repulsive}) are easily generalized to the case of mixtures  \cite{Cai-JSP,Cai-Mol1} or molecular fluids. Secondly,
when the pair interaction $w$ is neither attractive nor repulsive,  it is necessary to introduce two
real scalar fields $\varphi_{+} $ and $\varphi_{-} $ if some rigor is aimed at \cite{Cai-Mol}. Alternatively, in a more sloppy way, eq.~(\ref{attractive}) can be considered to hold in any cases having in mind that $\varphi$ will be a complex scalar field in the general case. We will thus  write formally in all cases
\begin{equation}
\label{csiKSSHE}
\Xi\left[ \nu \right]=\mathcal{N}_{w}^{-1} \int \mathcal{D} \varphi \;
\exp \left( - \mathcal{H}_{\mathrm{K}} \left[\nu, \varphi \right] \right) \; ,
\end{equation}
where the action (or effective Hamiltonian)  of the KSSHE field theory reads as
\begin{equation}
\label{action-K}
\mathcal{H}_K \left[\nu, \varphi \right]= \frac{1}{2}\left\langle \varphi \vert w^{-1} \vert \varphi \right\rangle -
\ln \Xi_{\mathrm{HS}}\left[ \overline{\nu} +  \varphi\right] \; .
\end{equation}
\subsection{The collective variables representation}
\label{CV}
We introduce now briefly the CV representation of $\Xi\left[ \nu \right]$ and refer the reader to a vast literature for a more
detailed presentation (see, e.g. \cite{Yuk1,Yuk-Hol,Yuk,Yuk2,Yuk3}). The starting point is the formally trivial identity
\begin{equation}
\label{a}
\exp \left(
\frac{1}{2}\left\langle \widehat{\rho}\vert w \vert \widehat{\rho}\right\rangle
\right) =\int \mathcal{D} \rho \;
 \delta_{\mathcal{F}}\left[ \rho -\widehat{\rho} \right]
 \exp \left(
\frac{1}{2}\left\langle \rho \vert w \vert \rho \right\rangle
\right) \; ,
\end{equation}
where $\delta_{\mathcal{F}}\left[ \rho \right]$ denotes the functional "delta"
\cite{Orland}. Making use  of its functional integral representation (see the appendix)
\begin{equation}
\delta_{\mathcal{F}}\left[ \lambda\right] \equiv
 \int \mathcal{D} \omega \; \exp \left(i \left\langle \omega
 \vert \lambda  \right\rangle \right) \;,
\end{equation}
one finds for the GC Boltzmann factor
\begin{equation}
\label{aa}
\exp \left(
\frac{1}{2}\left\langle \widehat{\rho}\vert w \vert \widehat{\rho}\right\rangle \right)=
 \int \mathcal{D} \rho  \mathcal{D} \omega \;
 \exp \left(  \frac{1}{2}\left\langle \rho \vert w \vert \rho \right\rangle
 +i \left\langle \omega \vert \left\lbrace
 \rho - \widehat{\rho}
  \right\rbrace \right\rangle
 \right) \; .
\end{equation}
Inserting equation~(\ref{aa}) in the definition~(\ref{csi}) of the GC partition function $ \Xi\left[ \nu \right]$
one obtains
\begin{equation}
\label{csiJac} \Xi\left[ \nu \right]= \int \mathcal{D} \rho  \; \exp \left( \frac{1}{2}
\left\langle \rho \vert w \vert \rho \right\rangle \right) \mathcal{J}\left[\rho,\overline{\nu} \right]   \; ,
\end{equation}
where the Jacobian 
\begin{equation}
\mathcal{J}\left[\rho,\overline{\nu} \right] =\int \mathcal{D} \omega \;
\exp \left(  i \left\langle \omega \vert \rho\right\rangle     \right) \; \Xi_{\mathrm{HS}}\left[ \overline{\nu} - i   \omega \right] 
\end{equation}
allows for the passage from the microscopic  variables $x_i$  to the collective ones $\rho$. We note that $\mathcal{J}\left[\rho,\overline{\nu} \right]$ does not depend on the pair interactions $w(1,2)$ but only
on the GC partition function of the reference HS system $\Xi_{\mathrm{HS}}\left[ \overline{\nu}\right] $ which is supposed
to be known.

Equation~(\ref{csiJac}) can also easily be recast under the form of a standard  statistical field theory, i.e. as
\begin{equation}
\label{csiCV} \Xi\left[ \nu \right]= \int \mathcal{D} \rho
 \mathcal{D} \omega \;
\exp \left( - \mathcal{H}_{\mathrm{CV}}\left[\nu, \rho, \omega \right] \right) \; ,
\end{equation}
where the action of the CV field theory reads as
\begin{equation}
\label{actionCV}
\mathcal{H}_{\mathrm{CV}} \left[\nu, \rho, \omega \right]= -\frac{1}{2}
\left\langle \rho \vert w \vert \rho \right\rangle  - i \left\langle \omega \vert \rho\right\rangle -
\ln \Xi_{\mathrm{HS}}\left[ \overline{\nu} - i   \omega \right] \; .
\end{equation}
We stress that $\omega$ and $\rho$ are two real scalar fields  and that equations~(\ref{csiCV}) and~(\ref{actionCV}) are valid for repulsive, attractive as well as  arbitrary pair interactions.
Moreover, with the clever normalization of Wegner \cite{Wegner} for the functional measures there are no unspecified
multiplicative constant involved in equation~(\ref{csiCV}) (see the appendix  for more details).

The CV transformation is clearly  more general than the KSSHE transformation since it  can be used for a pair interaction
$w(1,2)$ which does not possess an inverse and is easily generalized for n-body interactions ($n>2$). The equivalence of the CV and KSSHE representations~(\ref{csiKSSHE})
and~(\ref{csiCV}) of $\Xi\left[\nu \right]$ is readily established in the repulsive case ( $\widetilde{w}(q)<0$) by making use of the properties of Gaussian integrals (cf. equation~(\ref{totoi})). In the
attractive or in the general case we cannot propose a convincing way (i.e. a non formal one)
to establish this equivalence however.

\subsection{Statistical average}
\label{StatAver}
In the sequel it will be important to distinguish carefully, besides the usual GC average
$<\mathcal{A}(\mathcal{C})>_{\mathrm{GC}}$ of a dynamic variable $\mathcal{A}(\mathcal{C})$,
between two types of statistical field averages. At first the KSSHE averages defined as
\begin{equation}
\label{moyK}
\left\langle \mathcal{A}\left[\varphi \right] \right\rangle_{\mathrm{K}}= \Xi
\left[\nu \right]^{-1} \; \int \mathcal{D} \varphi \; \mathcal{A}\left[\varphi \right]
\exp \left( - \mathcal{H}_{\mathrm{K}} \left[\nu, \varphi \right] \right),
\end{equation}
where $\mathcal{A}\left[ \varphi \right]$ is some functional of the KSSHE fields $\varphi$ and, secondly the CV averages defined in a similar way as
\begin{equation}
\label{moyCV} \left\langle \mathcal{A}\left[\rho, \omega \right]
\right\rangle_{\mathrm{CV}}=\Xi\left[\nu \right]^{-1} \; \int
\mathcal{D} \rho
 \mathcal{D} \omega \; \; \mathcal{A}\left[\rho, \omega \right]
\exp \left( - \mathcal{H}_{\mathrm{CV}} \left[\nu, \rho, \omega \right] \right),
\end{equation}
where $\mathcal{A}\left[ \rho, \omega \right]$ is an arbitrary functional of the two CV fields  $\rho$ an $\omega $.
\section{Correlation functions}
\label{Corre} 
To establish the link between the usual theory of liquids and the two statistical field theories which were introduced in  sections~(\ref{HS}) and  ~(\ref{CV}) means to find the relations between  density correlation functions in the one hand and field (either KSSHE or CV) correlation functions on the other hand. This is the purpose of the present section.

\subsection{Density correlations}
\label{CorreGC}
The ordinary and truncated (or connected)  density correlation functions of the fluid will be defined in this paper as
\cite{Hansen,Stell1,Stell2}
\begin{eqnarray}
\label{defcorre}
G^{(n)}[\nu](1, \ldots, n) &=&\left< \prod_{1=1}^{n} \widehat{\rho}
    (x_{i}  \vert \mathcal{C}) \right>_{\mathrm{GC}} \; \nonumber \\
&=& \frac{1}{\Xi[\nu]}\frac{\delta^{n} \;\Xi[\nu]}
{\delta \nu(1) \ldots \delta \nu(n)}           \; ,\nonumber \\
G^{(n), T}[\nu](1, \ldots, n) &=&  \frac{\delta^{n} \log \Xi[\nu]}
{\delta \nu(1) \ldots \delta \nu(n)} \; .
\end{eqnarray}
Our notation emphasizes the fact that the  $G^{(n)}$ (connected and not connected) are functionals of the local chemical potential $\nu(x)$ and ordinary functions of the space coordinates $(1,\ldots, n) \equiv (x_{1},\ldots, x_{n})$.
We know from the theory of liquids that
\cite{Stell1,Stell2}
\begin{equation}
\label{Trunc}
G^{(n), T}[\nu](1,\ldots,n)= G^{(n)}[\nu]( 1,\ldots,n)
- \sum \prod_{m<n}G^{(m), T} [\nu](i_{1},\ldots,i_{m})  \; ,
\end{equation}
where the sum of products is carried out over all possible partitions of the set $(1,\ldots,n)$ into subsets of cardinal $m<n$. Of course $\rho[\nu](x) \equiv G^{(n=1)}[\nu](x)=G^{(n=1), T}[\nu](x)$ is the local density of the fluid.

It follows from the definition~(\ref{defcorre}) of the $G^{(n)}[\nu](1, \ldots, n)$ that they can be reexpressed as  KSSHE or  CV statistical averages, i.e.
\begin{eqnarray}
G^{(n)}[\nu](1, \ldots, n)&=&\left\langle
G^{(n)}_{\mathrm{HS}}[\overline{\nu} + \varphi](1, \ldots,
n)\right\rangle_{\mathrm{K}} \; ,  \\ G^{(n)}[\nu](1, \ldots,
n)&=&\left\langle G^{(n)}_{\mathrm{HS}}[\overline{\nu}  -
i\omega](1, \ldots, n)\right\rangle_{\mathrm{CV}} \; .
\end{eqnarray}
Although enlightening these relations are not very useful except for the special  case $n=1$ which reads explicitly as
\begin{eqnarray}
\rho\left[\nu \right](x)&=& \left\langle
\rho_{\mathrm{HS}}[\overline{\nu} +
\varphi](x)\right\rangle_{\mathrm{K}} \; , \\ \rho\left[\nu
\right](x)&=& \left\langle \rho_{\mathrm{HS}}[\overline{\nu}-
i\omega ](x)\right\rangle_{\mathrm{CV}} \; ,
\end{eqnarray}
where $\rho_{\mathrm{HS}}[\xi](x)$ is the local density of the hard sphere fluid at point $x$ in the presence of the local chemical potential $\xi(x)$.

\subsection{Field correlations}
\label{CorreK}
The correlation functions of the KSSHE field $\varphi$ and the CV fields $\rho$ and $\omega$ will of course will be defined as 
\begin{eqnarray}
\label{G-CVK}
G^{(n)}_{\varphi}[\nu](1, \ldots, n) &=&\left< \prod_{1=1}^{n} \varphi \left(x_{i}\right) \right>_{\mathrm{K}} \; , \nonumber \\
 G^{(n)}_{\rho}[\nu](1, \ldots, n) &=&\left< \prod_{1=1}^{n} \rho \left(x_{i}\right) \right>_{\mathrm{CV}} \; , \nonumber \\
 G^{(n)}_{\omega}[\nu](1, \ldots, n) &=&\left< \prod_{1=1}^{n} \omega \left(x_{i}\right) \right>_{\mathrm{CV}} \; ,
\end{eqnarray}
and their truncated counterparts will be defined in the same way as in equation~(\ref{Trunc}). The relations between the field correlation functions~(\ref{G-CVK}) and the density correlation functions  $G^{(n)}[\nu](1, \ldots, n)$ are easily obtained by introducing \emph{ad hoc} generating functionals and we quote here only the results, referring the reader to references~\cite{Cai-Mol,CPM} for a detailed dicussion.
\begin{itemize}
\item i) relations between  $ G^{(n)}_{\varphi}$ and  $G^{(n)}$

One has
\begin{eqnarray}
\label{dens-K}
\left\langle \varphi (1) \right\rangle_{\mathrm{K}} &=&w (1,1^{'})  \rho\left[ \nu \right] (1^{'})  \; ,
\nonumber \\
\label{dens-K2}
 G^{(2), T}_{\varphi}\left[ \nu \right] (1,2) &=& w (1,2) + w (1,1^{'}) w (2,2^{'}) G^{(2), T}\left[ \nu \right] (1^{'},2^{'})
\; , \nonumber \\
 \label{dens-Kn}
 G^{(n), T}_{\varphi}\left[ \nu \right] (1,\ldots,n) &=&
 w(1,1^{'})  \ldots w(n,n^{'})  \times \nonumber \\
 &\times &G^{(n), T}\left[ \nu \right] (1{'},\ldots,n{'}) \; 
 \mathrm{ for } \; \; n\geq 3 \; ,
\end{eqnarray}
where  we have adopted Einstein's convention, i.e.  space integration  of variables labeled by the same dummy indice  is meant. It will be quite enlighting to consider the GC correlations of the microscopic potential
\begin{equation}
\widehat{V}(x,\mathcal{C})\equiv \sum_{i=1}^{N}w(x,x_{i})=
 \int_{\Omega}d^{3}x^{'} \; \widehat{\rho}(x{'},\mathcal{C})) w(x,x{'})
 \;.
\end{equation}
We deduce readily  from equations~(\ref{dens-K}) that
\begin{eqnarray}
\left\langle \varphi (1) \right\rangle_{\mathrm{K}} &=&\left\langle \widehat{V}(1)\right\rangle _{\mathrm{GC}}  \; ,
\nonumber \\
G^{(2), T}_{\varphi}\left[ \nu \right] (1,2) &=&\left\langle \widehat{V}(1) \widehat{V}(2)\right\rangle^{T} _{\mathrm{GC}} 
 + w(1,2) \; , \nonumber \\
G^{(n), T}_{\varphi}\left[ \nu \right] (1,\ldots,n) &=&\left\langle \widehat{V}(1) \ldots \widehat{V}(n)\right\rangle^{T} _{\mathrm{GC}}  \mathrm{ for } \; \;  n\geq 3 \; .
\end{eqnarray}
Therefore the KSSHE field $\varphi(x)$ identifies "nearly" with the microscopic field $\widehat{V}(x)$ at a neutral point. 
\item ii) relations between $ G^{(n)}_{\rho}$, $ G^{(n)}_{\omega}$ and $G^{(n)}$ 

One finds easily that
\begin{equation}
\label{ru1}
G^{(n), T}_{\rho}\left[ \nu \right] (1,\ldots,n)=G^{(n), T}\left[ \nu \right] (1,\ldots,n) \; ,
\end{equation}
which was expected and that 
\begin{equation}
\label{ru2}
G^{(n), T}_{\varphi}\left[ \nu \right] (1,\ldots,n)=
  (-i)^n \; G^{(n), T}_{\omega}\left[ \nu \right] (1,\ldots,n) \;,
\end{equation}
which was suspected. We stress that the above relations are valid for all $n$. In particular, for $n=1$ note that 
$\left\langle \rho(x)\right\rangle _{\mathrm{CV}}=\rho(x)$ and $\left\langle \omega (x) \right\rangle _{\mathrm{CV}}= \mathrm{i}
\left\langle \widehat{V}(x)\right\rangle _{\mathrm{GC}} $. The CV $\rho$ and $\omega$ are thus connected to the density of particles and the density of energy respectively.
 We have therefore the correspondance $\varphi \leftrightarrow -\mathrm{i} \omega$ which will be a guideline in the sequel.
\end{itemize}

\section{Mean Field theory}
\label{MF}
Contrary to the situation which prevails in the theory of liquid, the term "mean-field" approximation is defined unambiguously in statistical field theory. It consists in neglecting field fluctuations and in approximating the grand potential  (i.e. the logarithm of the partition function) by minus the action at its saddle point \cite{Zinn}. In this section the KSSHE and CV theories will be considered at this level of approximation.
\subsection{KSSHE representation}
\label{MF-KSSHE}
The mean-field (MF) or saddle point approximation of the KSSHE theory will therefore be  defined by the set of relations \cite{Cai-Mol,Cai-JSP}:
\begin{equation}
\Xi_{\mathrm{MF}}\left[\nu \right]=\exp \left(- \mathcal{H}_{\mathrm{K}}\left[
\nu, \varphi_{0} \right]  \right) \; ,
\end{equation}
where, for $\varphi=\varphi_{0}$, the action is stationary, i.e.
\begin{equation}
\label{statio-K}\left. \frac{\delta \; \mathcal{H}_{\mathrm{K}}\left[
\nu, \varphi \right]}{\delta \varphi}\right \vert_{\varphi_{0}}=0 \; .
\end{equation}
Replacing the KSSHE action by its expression~(\ref{action-K}) in eq.~(\ref{statio-K})
leads to an  implicit equation for $\varphi_{0}$:
\begin{equation}
\label{statio2-K}
\varphi_{0}(1)=w(1,1^{'}) \; \rho_{\mathrm{HS}}\left[\overline{\nu} +
\varphi_{0}\right]( 1^{'}) \; ,
\end{equation}
which reduces to
\begin{equation}
\label{statio2-K-hom}
\varphi_{0}=\widetilde{w}(0)\; \rho_{\mathrm{HS}}\left[\overline{\nu} +
\varphi_{0}\right] \;
\end{equation}
for a homogeneous system.
It follows from the stationary condition~(\ref{statio-K}) that the MF density is given by
\begin{equation}
\label{ro-MF}
\rho_{\mathrm{MF}}\left[\nu \right] (1)= \frac{\delta \ln \Xi_{\mathrm{MF}}\left[\nu \right]}{\delta \nu(1)}=
\rho_{\mathrm{HS}} \left[ \overline{\nu} +
\varphi_{0} \right](1) \; ,
\end{equation}
and that the MF grand potential reads
\begin{equation}
\label{MF-gpot}
\ln \Xi_{\mathrm{MF}}\left[\nu \right]=
\ln \Xi_{\mathrm{HS}}\left[\overline{\nu} +
\varphi_{0}\right] -  \frac{1}{2}\left\langle \rho_{\mathrm{MF}}\vert w \vert\rho_{\mathrm{MF}}\right\rangle \; .
\end{equation}
Moreover, the MF Kohn-Scham free energy defined as the Legendre transform
\begin{equation}
\beta \mathcal{A}_{\mathrm{MF}}\left[\rho \right]  =\sup_{\nu}\left\lbrace
 \left\langle \rho \vert \nu\right\rangle -\ln \Xi_{\mathrm{MF}}\left[\nu \right]
 \right\rbrace
\end{equation}
is found to be the convex envelop of
\begin{equation}
\label{MF-A}
\beta \mathcal{A}_{\mathrm{MF}}\left[\rho \right]  =
\beta \mathcal{A}_{\mathrm{HS}}\left[\rho \right]
-\frac{1}{2}\left\langle\rho \vert w \vert \rho \right\rangle +\frac{1}{2} \int_{\Omega} dx \; w(0) \rho(x) \; .
\end{equation}
It can be shown \cite{Cai-Mol} that $\mathcal{A}_{\mathrm{MF}}\left[\rho \right]$
constitutes a rigorous upper bound for the exact free energy
$\mathcal{A}\left[\rho \right]$  if the interaction is attractive
($\widetilde{w}(q)>0$) and a lower bound in the converse case ($\widetilde{w}(q)<0$).

Finally, the pair correlation  and vertex (or direct correlation) functions at the
zero-loop order which are defined respectively as
\begin{eqnarray}
G_{\mathrm{MF}}^{(2)}\left[\nu \right](1,2)&=& \frac{\delta^{2} \ln \Xi_{\mathrm{\mathrm{MF}}}\left[ \nu\right]}
{\delta \nu(1) \; \delta \nu(2)} \; , \\
C_{\mathrm{MF}}^{(2)}\left[\rho \right] (1,2)&=& -\frac{\delta^{2}
\beta \mathcal{A}_{\mathrm{MF}}\left[\rho\right]} {\delta
\rho(1) \; \delta \rho(2)} \; ,
\end{eqnarray}
are given by
\begin{eqnarray}
G_{\mathrm{MF}}^{(2)}(1,2)&=& \left(1 -w \star
G^{(2)}_{\mathrm{HS}}\left[\overline{\nu} + \varphi_{0} \right]
\right)^{-1} \star  G^{(2)}_{\mathrm{HS}}\left[\overline{\nu} +
\varphi_{0} \right]\left(1,2 \right)   \\
\label{zozo}
C_{\mathrm{MF}}^{(2)}(1,2)&=&-G_{\mathrm{MF}}^{-1}(1,2) = C_{\mathrm{HS}}(1,2) + w(1,2) \; .
\end{eqnarray}
We note that these relations coincide with the old random phase approximation (RPA) of the theory of liquids \cite{Hansen}. Taking advantage of the arbitrariness of $w(r)$ in the core on can chose a regularisation of the potential such that the pair correlation function $g(r)=0$ for $r<\sigma$, then we recover the ORPA (optimized RPA) of the theory of liquids \cite{Hansen}. To be complete let us point out that   
it follows  from equation~(\ref{dens-K2}) that the free propagator of the KSSHE field theory is given by
\begin{equation}
\label{propa} G_{\varphi, \;\mathrm{MF}}^{(2)}(1,2)= \left(1 -w
\star G^{(2)}_{\mathrm{HS}}\left[\overline{\nu} + \varphi_{0}
\right](1,2)\right)^{-1}
 \star w(1,2) \; .
\end{equation}

\subsection{CV representation}
\label{MF-CV}
An analysis similar to that of section~(\ref{MF-KSSHE}) can be made in the CV representation.
The MF level of the CV field theory will be defined by
\begin{equation}
\Xi_{\mathrm{MF}}\left[\nu \right]=\exp \left(- \mathcal{H}_{\mathrm{CV}}\left[
\nu, \rho_{0}, \omega_{0} \right]  \right) \; ,
\end{equation}
where, for $\rho=\rho_{0}$ and $\omega=\omega_{0}$ the CV action is stationary, i.e.
\begin{equation}
\label{statio-CV}\left. \frac{\delta \; \mathcal{H}_{\mathrm{CV}}\left[
\nu, \rho, \omega \right]}{\delta \rho}\right \vert_{(\rho_0,\omega_{0})}
=\left. \frac{\delta \; \mathcal{H}_{\mathrm{CV}}\left[
\nu, \rho, \omega \right]}{\delta \omega}\right \vert_{(\rho_0,\omega_{0})}
=0 \; .
\end{equation}
Replacing the CV action by its expression~(\ref{actionCV}) in
equation~(\ref{statio-CV}) yields a set of two coupled implicit
equations for $\rho_{0}$ and $\omega_{0}$:
\begin{eqnarray}
0&=& w(1,2)\rho_{0}(2) + i \omega_{0}(1) \nonumber \; , \\
0&=&  \rho_{0}(1) - \rho_{\mathrm{HS}}\left[ \overline{\nu} -i\omega_{0} \right](1) \; .
\end{eqnarray}
If we define $\varphi_{0}= -i \omega_{0}$, then the two previous equations can be rewritten
\begin{eqnarray}
\varphi_{0}&=& w(1,2) \rho_{0}(2) \nonumber \; , \\
 \rho_{0}(1)&=& \rho_{\mathrm{HS}}\left[ \overline{\nu} + \varphi_{0}\right](1) \; ,
\end{eqnarray}
which shows that, as expected, $\varphi_{0}$ coincides with the saddle point of the
KSSHE field theory (cf  Sec~\ref{MF-KSSHE}). Moreover a direct calculation will show that
\begin{equation}
\ln \Xi_{\mathrm{MF}}\left[\nu \right]= -\mathcal{H}_{\mathrm{K}}\left[\nu, \varphi_{0} \right]
= -\mathcal{H}_{\mathrm{CV}}\left[\nu, \rho_{0},\omega_{0} \right] \; .
\end{equation}
Therefore the local density, the grand potential, the  free energy, the  correlation and the direct correlation functions coincide at the MF level in the CV and KSSHE field theories.
\section{Loop expansion}
\label{loop}
The loop expansion is a usefull tool of statistical field theory \cite{Zinn}. Away from the critical point it yields approximations for the free energy beyond the MF level.
\subsection{Loop expansion of the grand potential}
\label{loopp}
The loop expansion of the logarithm of the partition function of a scalar field theory can be found in any standard text book, see e.g. that of Zinn-Justin \cite{Zinn}. In the case of the KSSHE field theory we have just to reproduce known results. One proceeds as follows. A small dimensionless parameter $\lambda$ is introduced and the loop expansion is obtained by the set of transformations
\begin{eqnarray}
\varphi &=&\varphi_{0} + \lambda^{1/2} \chi \; ,\nonumber \\
\ln \Xi\left[ \nu \right] &=& \lambda \ln \left\lbrace
\mathcal{N}_{w}^{-1} \int \mathcal{D}\chi \;
\exp \left( - \frac{\mathcal{H}_{\mathrm{K}}\left[\nu,\varphi \right]}{\lambda} \right)
\right\rbrace \; , \nonumber \\
 &=&\ln \Xi^{(0)}\left[ \nu \right] + \lambda \ln \Xi^{(1)}\left[ \nu \right]+
 \lambda^{2} \ln \Xi^{(2)}\left[ \nu \right] + \mathcal{O}(\lambda^{3}) \; ,
\end{eqnarray}
where $\varphi_{0}$ is the saddle point of the KSSHE action. The expansion of $\ln \Xi\left[ \nu \right]$ in powers of $\lambda$ is obtained by performing a  cumulant expansion and by making a repeated use  of Wick's theorem. At the end of the calculation one set $\lambda=1$.

In order to obtain a loop expansion of $\ln \Xi$ in the CV representation a little homework is necessary.
In this case we are led to introduce the following set of transformations
\begin{eqnarray}
\label{aaa}\rho &=&\rho_{0} + \lambda^{1/2} \delta \rho  \; , \\
\label{bb}\omega &=&\omega_{0} + \lambda^{1/2} \delta \omega \; , \\
\label{cc}\ln \Xi\left[ \nu \right] &=& \lambda \ln \left\lbrace  \int \mathcal{D}\delta\rho \;
\mathcal{D}\delta\omega \;
\exp \left( - \frac{\mathcal{H}_{\mathrm{CV}}\left[\nu,\rho, \omega \right]}{\lambda} \right)
\right\rbrace \; ,
\end{eqnarray}
where $(\rho_{0},\omega_{0})$ is the saddle point of the CV action. The form retained in equations~(\ref{aaa}) and~(\ref{bb}) is imposed by the exact relation $\left\langle \omega \right\rangle_{\mathrm{CV}}(1)= \mathrm{i} w(1,1^{'}) \left\langle \rho
\right\rangle_{\mathrm{CV}} (1^{'})$  (cf. equations~(\ref{ru1})  and~(\ref{ru2})  for $n=1$). Then one performs a cumulant expansion of~(\ref{cc}) and use the appropriate version of Wick's theorem (i.e. for two coupled Gaussian  fields $\rho$ and $\omega$). The derivation is detailed in reference~\cite{CPM} and will not be reproduced here. One finds that the KSSHE and CV loop expansions of $ \ln \Xi\left[ \nu \right]$ coincide at each order of the expansion (i.e. the $\ln \Xi^{(n)}\left[ \nu \right] $ are the same) which was expected.

\begin{figure}[!ht]
\begin{center}
\epsfig{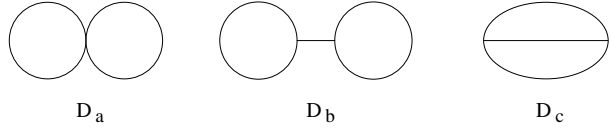}
\caption{\label{diag}
Diagrams which contribute to $ \ln \Xi^{(2)}\left[ \nu \right]$. $D_a$ and $D_c$ are irreducible while $D_b$ is reducible.}
\end{center}
\end{figure}

Of course $ \Xi^{(0)}\left[ \nu \right]\equiv  \Xi_{\mathrm{MF}}\left[ \nu \right]$. At the one-loop order one finds \cite{Cai-Mol,CPM}
\begin{equation}
\label{Xi1-K} \Xi^{(1)}\left[ \nu \right]=
\frac{\mathcal{N}_{\Delta_{\varphi_{0}}}}{\mathcal{N}_{w}}=\frac{\int
\mathcal{D}\varphi \; \exp \left( -\frac{1}{2}\left \langle
\varphi \vert \Delta_{\varphi_{0}}\vert  \varphi \right \rangle
\right)}{\int \mathcal{D}\varphi \; \exp \left( -\frac{1}{2}\left
\langle \varphi \vert w \vert \varphi \right \rangle \right)  } \; ,
\end{equation}
where we have adopted Zinn-Justin's notations for the free propagator \cite{Zinn}
 \begin{equation}
\label{propa-2} \Delta_{\varphi_{0}}^{-1}(1,2)=
w^{-1}(1,2) -G_{\mathrm{HS}}^{(2), \; T}\left[ \overline{\nu} +\varphi_{0}\right] (1,2)
\; .
\end{equation}
Of course  $\Delta_{\varphi_{0}}(1,2)$  coincides with $G_{\varphi, \;\mathrm{MF}}^{(2)}(1,2)$  as can be seen by comparing equations~(\ref{propa}) and~(\ref{propa-2}). If $\nu$ is uniform then the system is homogeneous and $\Delta_{\varphi_{0}}(1,2)$ takes on a simple form in Fourier space, i.e.
\begin{equation}
 \widetilde{\Delta}_{\varphi_{0}}(q)=\frac{\widetilde{w}(q)}
{1- \widetilde{w}(q) \widetilde{G}_{\mathrm{HS}}^{(2), \; T}\left[ \overline{\nu} +\varphi_{0}\right](q) } \; ,
\end{equation}
and, in this case, the Gaussian integrals in equation~(\ref{Xi1-K}) can be performed explicitly  (cf the appendix for more details). One finds that
\begin{equation}
\label{Xi1-K-bis}
\ln \Xi^{(1)}\left[ \nu \right]= -\frac{V}{2}
\int_{q} \ln \left(1 - \widetilde{w}(q)   \widetilde{G}_{\mathrm{HS}}^{(2), \; T}\left[ \overline{\nu}
+\varphi_{0} \right] (q)  \right) \; .
\end{equation}
The second-loop order contribution $\ln \Xi^{(2)}$ to the  grand
potential has a complicated expression involving the sum of three
diagrams sketched in figure.~\ref{diag}
\begin{equation}
\label{Xi2-K}
\ln \Xi^{(2)}\left[ \nu \right]= D_a +D_b +D_c \; .
\end{equation}

More explicitly one has \cite{Zinn}
\begin{eqnarray}
\label{D}
D_a&=&
\frac{1}{8} \int_{\Omega} d1 \ldots d4 \;
\Delta_{\varphi_{0}}(1,2) \Delta_{\varphi_{0}}(3,4)
G_{\mathrm{HS}}^{(4), \; T}\left[ \overline{\nu} +\varphi_{0} \right] (1,2,3,4) \nonumber \\
D_b&=&
\frac{1}{8} \int_{\Omega} d1 \ldots d3 \;  d1^{'} \ldots d3^{'}\;
\Delta_{\varphi_{0}}(1,2)  \Delta_{\varphi_{0}}(1^{'},2^{'}) \Delta_{\varphi_{0}}(3,3^{'})    \nonumber \\
& \times & G_{\mathrm{HS}}^{(3), \; T}\left[ \overline{\nu} +\varphi_{0} \right] (1,2,3)
G_{\mathrm{HS}}^{(3), \; T}\left[ \overline{\nu} +\varphi_{0} \right] (1^{'},2^{'},3^{'}) \nonumber \\
D_c&=&
\frac{1}{12} \int_{\Omega} d1 \ldots d3 \;  d1^{'} \ldots d3^{'}\;
\Delta_{\varphi_{0}}(1,1^{'}) \Delta_{\varphi_{0}}(2,2^{'}) \Delta_{\varphi_{0}}(3,3^{'}) \nonumber \\
& \times &
 G_{\mathrm{HS}}^{(3), \; T}\left[ \overline{\nu} +\varphi_{0} \right] (1,2,3)
G_{\mathrm{HS}}^{(3), \; T}\left[ \overline{\nu} +\varphi_{0} \right] (1^{'},2^{'},3^{'}) \; .
\end{eqnarray}
As they stand, the above relations are not particularly  useful for practical applications (even in the homogeneous case) since they involve the three and four body  correlation functions of the reference HS fluid. We will introduce some reasonable approximation in section~(\ref{free}) to tackle with this horrible expression. Quite remarkably it has been shown recently that for a symmetric mixture of charged hard spheres $\ln \Xi^{(2)}$ has a much more simple expression which involves only the pair correlation functions of the HS fluid as a consequence of local charge neutrality \cite{Cai-JSP,Cai-Mol1}.
\subsection{Loop expansion of the pressure and the free energy}
\label{free}
In this section we restrict ourselves to the homogeneous case, therefore $\ln \Xi\left[\nu \right]=V \beta P\left(\nu \right)$,
where $P$ denotes the pressure and $\beta \mathcal{A}\left[\rho \right]= V \beta f(\rho)$ where $f$ is the Helmoltz free energy per unit volume. The two-loop expression that we derived for $P$ in section~(\ref{loopp}) is too much complicated to be of any practical use  since it involves the $3$ and $4$ body density correlation functions of the HS fluid which are unknown, whereas $G_{\mathrm{HS}}^{(2), \; T}$ is known approximatively, for instance
in the Percus-Yevick (PY) approximation \cite{Hansen}. A simple but coherent approximations for $G_{\mathrm{HS}}^{(3), \; T}$ and $G_{\mathrm{HS}}^{(4), \; T}$ will be proposed now.

Recall first that it follows from their definitions \cite{Stell1,Stell2} (see e.g. equations~(\ref{defcorre}))
that the $G_{\mathrm{HS}}^{(n), \; T}\left[\nu \right] $ satisfy to the following relations
\begin{equation}
\label{h1}\frac{\delta }{\delta \nu(n+1)}
G_{\mathrm{HS}}^{(n), \; T}\left[\nu \right](1, \ldots,n)=G_{\mathrm{HS}}^{(n+1), \; T}\left[\nu \right](1, \ldots,n,n+1) \; .
\end{equation}
For a homogeneous system (in which case  $\nu$ is a constant) one infers from this equation that
\begin{equation}
\label{h2}\int_{\Omega}d1 \ldots dn \; G_{\mathrm{HS}}^{(n+1), \; T}\left[\nu \right](1, \ldots,n,n+1)=
\frac{\partial^{n}}{\partial \nu^{n}}\rho_{\mathrm{HS}}\left(\nu \right)\equiv
\rho_{\mathrm{HS}}^{(n)}\left(\nu \right)
\; ,
\end{equation}
where $\rho_{\mathrm{HS}}\left(\nu \right)$ is the number density of hard spheres at the chemical potential $\nu$.
In the rest of the section we will adopt the following approximation
\begin{equation}
\label{hyp}
G_{\mathrm{HS}}^{(n), \; T}\left[\nu \right](1, \ldots,n)=
\rho_{\mathrm{HS}}^{(n+1)}\left(\nu \right) 
\delta(n,1) \ldots \delta(2,1)
\; \mathrm{for} \; \; n \geq 3 \; .
\end{equation}
Note that this hypothesis is coherent with the exact relations~(\ref{h1}) and~(\ref{h2}). However, we keep the full
$G_{\mathrm{HS}}^{(2), \; T}\left[\nu \right](1,2) \equiv G_{\mathrm{HS}}^{(2), \; T}\left[\nu \right](x_{12})$ with the assumption that it is  a known function of $x_{12}$  (in the PY approximation for instance). The free propagator $\Delta_{\varphi_{0}}(x_{12})$  has thus also a well defined expression.
It is not difficult to convince oneself that that the set of approximations that we have introcuced is reasonable as long as the range of the KSSHE field correlation
functions is (much) larger than than the range of the HS density correlation functions. This will  be true if $w$ is a long range pair interaction (near  the Kacs limit for instance).

With the hypothesis~(\ref{hyp}) it is easy to obtain the two-loop
order approximation for the pressure. One finds
\begin{eqnarray}
\label{press}\beta P (\nu) &=&\beta P^{(0)}(\nu) + \lambda \beta P^{(1)}(\nu)+
\lambda^{2} \beta P^{(2)}(\nu)+\mathcal{O}(\lambda^{3}) \; ,\nonumber \\
\beta P^{(0)}(\nu)&=&\beta P_{\mathrm{MF}}(\nu)=P_{\mathrm{HS}}(\overline{\nu}+\varphi_{0})-\frac{\varphi_{0}^{2}}{
2 \widetilde{w}(0)} \; ,\nonumber \\
\beta P^{(1)}(\nu)&=&-\frac{1}{2} \int_{q} \ln\left(
1- \widetilde{w}(q)\widetilde{G}_{\mathrm{HS}}^{(2), \; T}\left[
\overline{\nu}+\varphi_{0}
\right](q) \right) \; ,\nonumber \\
\beta P^{(2)}(\nu)&=&\frac{\rho^{(3)}_{0}}{8}\Delta_{\varphi_{0}}^{2}(0)+
\frac{\left[ \rho^{(2)}_{0}\right] ^{2}}{8} \widetilde{\Delta}_{\varphi_{0}}(0) \Delta_{\varphi_{0}}^{2}(0) \nonumber \\
&+& \frac{\left[ \rho^{(2)}_{0}\right] ^{2}}{12}\int d^{3}x \; \Delta_{\varphi_{0}}^{3}(x) \; ,
\end{eqnarray}
where $\rho^{(n)}_{0} \equiv \rho_{\mathrm{HS}}^{(n)}\left( \overline{\nu}+\varphi_{0}\right)$ and
$\int_{q}\equiv \int d^{3}q/(2 \pi)^{3}$.

The free energy $\beta f(\rho)$ is obtained as the Legendre transform of $\beta P(\nu)$. The calculation is lengthy and tricky and we must contend ourselves to give the final result (see reference \cite{CPM} for the complete proof). Setting $\lambda=1$ one finds
\begin{eqnarray}
\label{f2l}
\beta f(\rho)&=& \beta f_{\mathrm{HS}}(\rho) -\frac{\widetilde{w}\left( 0\right)}{2}
\rho^{2} \nonumber \\
&+&\frac{1}{2} \int_{q} \left\lbrace \ln \left(1 -
 \widetilde{w}\left( q\right)  \widetilde{G}_{\mathrm{HS}, \; \rho}^{T}\left( q\right)
 \right)  + \rho \widetilde{w}\left( q\right)\right\rbrace  \nonumber \\
 &-&\frac{\rho^{(3)}}{8}\Delta_{\rho}^{2}(0)+
\frac{1}{8}  \Delta_{\rho}^{2}(0)\frac{\left[ \rho^{(2)}\right]^{2} }{\rho^{(1)}}
-  \frac{\left[ \rho^{(2)}\right] ^{2}}{12}\int d^{3}x \;
\Delta_{\rho}^{3}(x) \; ,
\end{eqnarray}
where the HS pair correlation function  $G_{\mathrm{HS}, \; \rho}^{T}$ must be evaluated at the density $\rho$ and, similarly,
\begin{equation}
\widetilde{\Delta}_{\rho}(q)=\frac{\widetilde{w}\left( q\right)}{1 -\widetilde{G}_{\mathrm{HS}, \; \rho}^{T}
\left( q\right)\widetilde{w}\left( q\right) } \; .
\end{equation}
Finally, in  equation~(\ref{f2l}) 
\begin{eqnarray}
\rho^{(1)}&=&\frac{1}{\nu_{\mathrm{HS}}^{(1)}\left( \rho\right) } \nonumber \\
\rho^{(2)}&=&\frac{\partial\rho^{(1)}_{0} }{\partial \nu} =\frac{-\nu_{\mathrm{HS}}^{(2)}(\rho)}
{\left[\nu_{\mathrm{HS}}^{(1)}(\rho)\right]^{3}} \nonumber \\
\rho^{(3)}&=&\frac{\partial\rho^{(2)}_{0} }{\partial \nu} =
\frac{3 \left[\nu_{\mathrm{HS}}^{(2)}(\rho)\right]^{2} -
\nu_{\mathrm{HS}}^{(3)}(\rho)\nu_{\mathrm{HS}}^{(1)}(\rho)
}{\left[\nu_{\mathrm{HS}}^{(1)}(\rho)\right]^{5}} \; ,
\end{eqnarray}
where $\nu_{\mathrm{HS}}^{(n)}(\rho)$ denotes the $n$th derivative of the HS chemical potential with respect to the density (it can be computed in the framework of the PY or Carnahan-Starling approximations for instance \cite{Hansen}).
Some remarks are in order.

i) It must be pointed out that, quite unexpectedly, the reducible diagram $D_b$ has not be cancelled by the Legendre
transform. Usually, in statistical field theory it is the case (cf. \cite{Zinn}). The reason is that  the chemical potential
$\nu$ is not the field conjugate to the order parameter $m=\left\langle \varphi \right\rangle_{\mathrm{K}}$ of the KSSHE field theory. However one of us  have shown elsewhere \cite{Cai-JSP,Cai-Mol1} that for the symmetric mixtures of charged hard spheres only irreducible diagrams contribute to $\beta f^{(2)}$.

ii) The two first terms of equation~(\ref{f2l}) (i.e. the one-loop order result) supplemented by equation~(\ref{propa}) for the pair correlation function are  \emph{exactly} constitutive of the RPA theory of liquid. The full result including the two-loop contribution is a new approximation as far as the authors know.

iii) All the quantities which enter equation~(\ref{f2l}) can be computed numerically (for instance in the PY approximation). However we stress once again that the validity of equation~(\ref{f2l}) is restricted to long range pair potentials (for instance $w(r)=\gamma^{3} \Phi(r/\gamma)$ with $\gamma\rightarrow 0$) and we are not aware of exact results (i.e. obtained by Monte Carlo simulations) which would allow us to test the validity of our expression for the free energy for such kind of fluids.
\section{Conclusion}
\label{conclu}
Using the CV method we have reconsidered the basic relations of the statistical field theory of simple fluids that follow from this approach. In contrary to the KSSHE theory \cite{Cai-Mol} the corresponding CV action depends on two scalar fields, the field $\rho$ connected to the number density of particles and the field $\omega$ conjugate to $\rho$, connected to the density of energy. Explicit relations between field and density correlation functions were obtained.

For a one-component continuous model of fluid, consisting of hard
spheres interacting through an attractive pair potential, we have calculated the grand partition function for both versions of statistical field theory using the loop expansion technique. As it was expected, at all the orders of loop expansion considered, both versions of the theory produced indeed the same analytical results. The expressions for the pressure as well as for the free energy were  derived at the two-loop level. It yields a new type of approximation and Monte Carlo simulations are wanted to test its validity.

From our analysis of the CV and KSSHE transformations we can also conclude that the former has some important advantages which could be very useful for more complicated models of fluids. In particular, it is valid for an arbitrary pair potential (including a pair interaction $w(1,2)$ which does not possess an inverse) and is easily generalized for the case of n-body interparticle interactions with $n>2$.

The statistical field theory of liquids is still in its enfancy and many progresses including the implementation of Wilson's ideas on the renormalization group for studying gas-liquid critical points are likely to be achieved in the future, this is at least the hope of the authors.

\section*{Acknowledgements}
This work was made in the framework of the cooperation project
between the CNRS and the NASU (ref.~ CNRS 17110).
\newpage
\appendix
\label{appendix}
\section*{Functional measures and integrals}
We give in this appendix some details concerning functional measures and integrals. Let us consider a real scalar field $\varphi(x)$ defined in a cube $\mathcal{C}_3$ of side $L$ and volume $V=L^3$. We assume periodic boundary conditions, i.e. we restrict ourselves to fields which can be written as a Fourier series,
\begin{equation}
\varphi(x)=\frac{1}{L^3} \; \sum_{q \in \Lambda}
\widetilde{\varphi}_{q} \; \mathrm{e}^{ \mathrm{i} q x} \;,
\end{equation}
where $\Lambda = (2 \pi/L)\;  \Z^3$ is the reciprocal cubic lattice. The reality of $\varphi$ implies that, for $q\ne
0$ $\widetilde{\varphi}_{q} =
\widetilde{\varphi}_{-q}^{\star}$, where the star means complex conjugation.
 Following Wegner \cite{Wegner} we define the normalized functional measure $\mathcal{D}\varphi$ as
\label{dphi}
\begin{eqnarray}
\mathcal{D} \varphi & \equiv & \prod_{q  \in \Lambda}
\frac{d \widetilde{\varphi}_{q} }
{\sqrt{2 \pi  V}} \\
d \widetilde{\varphi}_{q} d
\widetilde{\varphi}_{-q} & = & 2 \;
d \mathrm{Re}\widetilde{\varphi}_{q} \;
d \mathrm{Im}\widetilde{\varphi}_{q} \;   \mathrm{ for } \;  q  \ne 0 \; .
\end{eqnarray}
Eq.\ (\ref{dphi}) can thus  be conveniently rewritten as
\begin{equation}
\label{dphi_bis} \mathcal{D} \varphi= \frac{d \varphi_{0}}{\sqrt{2
\pi  V}} \prod_{q \in \Lambda^{\star}}
\frac{d \mathrm{Re} \widetilde{\varphi}_{q} \;d \mathrm{Im} \widetilde{\varphi}_{q}}{\pi V} \; ,
\end{equation}
where the sum in the right hand side  runs over only the half $\Lambda^{*}$ of all the vectors of the reciprocal lattice $\Lambda$ (for instance those with $q_x \geq 0$). The normalization constant of equation ~(\ref{dphi}) can thus be explicitely evaluated as
\begin{eqnarray}
\label{norma}
\mathcal{N}_{w}&=& \int \mathcal{D} \varphi \;  \exp \left( -\frac{1}{2}
\left\langle \varphi \vert w^{-1} \vert \varphi \right\rangle \right) \nonumber \\
&=& \exp \left(\frac{1}{2 } \sum_{{q} \in \Lambda}  \ln \widetilde{w}(q) \right)  
\sim \exp \left( \frac{V}{2}\;
\int_{q} \ln \widetilde{w}(q) \right) \; ,
\end{eqnarray}
where $w(1,2)$ is positive and satisfies $w(1,2)=w(2,1)\equiv
w(x_{12})$ and  $(2 \pi)^{3} \int_{q} \equiv \int d^{3}q $.  It is worth noting that in this case we have trivially
$\mathcal{N}_{w^{-1}}=1/\mathcal{N}_{w}$. More generally and with
the same hypothesis we have the two useful identities
\begin{eqnarray}
\label{Gauss}
\left\langle \exp \left( (i)\left\langle \varphi \vert  \omega \right\rangle \right)
\right\rangle_{w}
&\equiv&
\mathcal{N}_{w}^{-1} \int \mathcal{D} \varphi \;
\exp \left( -\frac{1}{2}
\left\langle \varphi \vert w^{-1} \vert \varphi \right\rangle + (i)
\left\langle \varphi \vert  \omega \right\rangle \right) \nonumber \\
&=&
\exp \left(
 +(-) \frac{1}{2}\left\langle \omega \vert w \vert \omega \right\rangle
\right) \; ,
\end{eqnarray}
where $\omega$ is a real scalar field.

We define now the "functional delta" distribution $\delta_{\mathcal{F}}\left[ \lambda \right] $  as
\begin{equation}
\label{deltaF}
\delta_{\mathcal{F}}\left[ \lambda\right] \equiv
 \int \mathcal{D} \omega \; \exp \left(i \left\langle \omega
 \vert \lambda  \right\rangle \right) \;,
\end{equation}
where both $\omega$ and $\lambda$ are \emph{real} scalar fields defined on $\mathcal{C}$. Since
\begin{equation}
\left\langle \omega
 \vert \lambda  \right\rangle=
\int_{\mathcal{C}} d^3 x \; \omega(x) \lambda (x)=\frac{1}{V}\sum_{q \in \Lambda}
 \widetilde{\omega}_{-q}\widetilde{\lambda}_{q} \; ,
\end{equation}
it follows from equation~(\ref{dphi}) that we have more explicitly
\begin{equation}
\label{deltaF-bis} \delta_{\mathcal{F}}\left[ \lambda\right]=
\sqrt{2 \pi V} \; \delta \left( \widetilde{\lambda}_0 \right)  \;
\prod_{q \in \Lambda^{\star}}\left[ \pi V \; \delta
\left( \mathrm{Re}\widetilde{\lambda}_{q} \right)  \; \delta
\left( \mathrm{Im}\widetilde{\lambda}_{q} \right) \right] \; .
\end{equation}
Therefore
\begin{equation}
\int \mathcal{D} \lambda \;\delta_{\mathcal{F}}\left[ \lambda\right]=1 \;,
\end{equation}
and, more generally
\begin{equation}
\int \mathcal{D} \lambda \;
\mathcal{F}\left[\lambda \right]
\delta_{\mathcal{F}}\left[ \lambda
- \lambda _0\right]=\mathcal{F}\left[\lambda_0 \right]
 \;,
\end{equation}
where  $\mathcal{F}\left[\lambda \right] $ is some arbitrary functional
of the field $\lambda(x)$.


%


%
%
%


%

\end{document}